\documentstyle[twocolumn,prl,aps,epsf,floats]{revtex}
\begin{document}
\draft
\def \beq{\begin{equation}}
\def \eeq{\end{equation}}
\def \beqarr{\begin{eqnarray}}
\def \eeqarr{\end{eqnarray}}

\twocolumn[\hsize\textwidth\columnwidth\hsize\csname @twocolumnfalse\endcsname

\title{Field Theoretical Description of
Quantum Hall Edge Reconstruction
}

\author{Kun Yang}

\address{
National High Magnetic Field Laboratory and Department of Physics,
Florida State University, Tallahassee, Florida 32306
}

\date{\today}
 
\maketitle

\begin{abstract}
We propose a generalization of the chiral Luttinger liquid theory to allow 
for a unified description of quantum Hall edges with or without edge 
reconstruction. Within this description edge reconstruction is found to be
a quantum phase transition 
in the universality class of
one-dimensional dilute Bose gas transition, whose critical behavior can be
obtained exactly. At principal filling factors $\nu=1/m$, we show
the additional edge modes
due to edge reconstruction 
modifies the point contact tunneling exponent in the 
low energy limit, by a small and non-universal amount.

\end{abstract}

]

Recently there has been considerable interest in the physics at the edge of a 
quantum Hall liquid\cite{review}. Our 
theoretical understanding of the
edge physics is mostly based on the chiral Luttinger liquid 
(CLL) theory advanced by Wen\cite{wen}.
The CLL theory is a long-wave length, low-energy effective field theory 
which is closely tied to the fundamental topological features of the bulk 
quantum Hall liquid, and describes the most robust physical properties of the
edges states, including the quantization of Hall conductance. It also makes 
a number of remarkable prediction about single-particle properties at the edge;
for example it predicts that in point contact tunneling between a Fermi liquid 
metal and a quantum Hall edge, the
current-voltage relation follows a power-law $I\sim V^\alpha$, which is 
characteristic of a Luttinger liquid,
and for a whole sequence of bulk filling factors, the exponent $\alpha$ is
{\em universal} and independent of the details of the edge confining potential
and electron-electron interaction. Such power law behavior has been
observed in recent tunneling experiments\cite{chang,grayson,chang1,hilke}, 
and at principal filling factors like $\nu=1/3$, the exponent $\alpha$ 
was found to be
close to but noticeably different from
the CLL prediction\cite{chang}. Away from $\nu=1/3$ 
however, more
significant discrepancy between theory and experiment
has been found\cite{grayson,chang1,hilke}.

In the meantime, {\em microscopic} theoretical studies have suggested that 
the interplay between electron-electron interaction and confining potential at
the edge at shorter distance (typically of order magnetic length $\ell$ or 
slightly above that)
can give rise to nontrivial low-energy 
physics. In particular, it was found that the competition between the two can
lead to edge reconstruction, both at integer\cite{macdonald,chamon}
and fractional\cite{wan,wry} bulk filling. In particular, it was argued 
recently\cite{wan,wry} that
for realistic sample parameters, edge reconstruction is essentially always 
present for fractional bulk filling, despite the presence of a
sharp potential barrier in the samples grown by the cleaved edge overgrowth
technique used in the recent tunneling experiments. 
Edge reconstruction gives rise to {\em additional} low-energy
edge modes\cite{chamon,wan,wry}
that are not described by the original
CLL theory\cite{note}; these additional modes can 
profoundly affect the low-energy physics at the edge. 

In this paper we propose a generalization of the CLL theory to accommodate the
short-distance physics. In our generalized theory the edge with and without
reconstruction can be described on equal footing; in particular, the 
edge reconstruction transition between these two different phases can be 
studied.
For simplicity and clarity we will focus on principal filling factors 
$\nu=1/m$ with $m$ being an odd integer; generalization to other filling factors
is conceptually straightforward.
We find that the edge reconstruction is a quantum phase transition 
in the universality class of
one-dimensional dilute Bose gas transition, whose critical behavior can be
obtained exactly. We also show that 
the additional edge modes
due to edge reconstruction 
modifies the power-law exponent of the single electron Green's
function, and thus the
point contact tunneling exponent in a non-universal way, although the 
modification is likely to be quantitatively rather small. 
We will also make contact with recent experiments and existing theories
on edge tunneling.

For principal bulk filling $\nu=1/m$, there is one chiral edge mode described
by the following Hamiltonian within the CLL theory\cite{wen}:
\beq
H=2\pi mv\sum_{k> 0}\rho_k\rho_{-k}=\pi mv\int{dx\rho^2(x)},
\label{cllh}
\eeq
where $v$ is the velocity for edge excitations, $\rho_k$ is the momentum space
edge electron density operator which satisfies the Kac-Moody algebra:
$[\rho_k, \rho_{k'}]=-{k\over 2\pi m}\delta_{k+k'}$,
and $\rho(x)$ is the corresponding
edge electron density operator in real space. Eq. (\ref{cllh})
describes a single branch of chiral bosons with {\em linear} dispersion:
$\epsilon_k=vk$,
{\em i.e.}, the bosons propagate with a fixed velocity $v$. 
In real space Eq. (\ref{cllh}) describes
a density-density coupling that is completely {\em local}.
While it is
appropriate to neglect the non-locality of the electron-electron interaction
in the long-wave length limit, we must also keep in mind that
what drives edge reconstruction is the interplay between
electron-electron interaction
and confining potential at shorter distances; the
typical length scale associated with edge reconstruction is the magnetic length
$\ell$. We thus generalize Eq. (\ref{cllh}) to incorporate the
non-local nature of the electron-electron interaction:
\beqarr
&H&= \int{dx}\int{dx'}\rho(x)V(x-x')\rho(x')\nonumber\\
&=&\pi mv\int{dx\{\rho^2(x)
-a[\partial_x\rho(x)]^2+b[\partial^2_x\rho(x)]^2+\cdots\}}.
\label{rch}
\eeqarr
In (\ref{rch}) we have performed a gradient expansion, and $a$ and $b$ are
phenomenological constants that depend on details of the electron-electron
interaction; for a generic short-range
repulsive interaction
we have $a > 0$ and $b > 0$, thus we can truncate
the gradient expansion to the corresponding order without losing stability of
the model.

\begin{figure}
\epsfxsize=3.6in
\centerline{ \epsffile{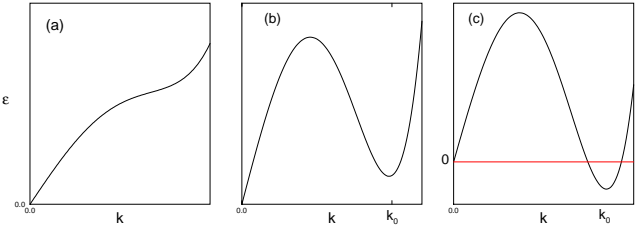} }
\caption{
Chiral boson dispersion for different situations. (a) $a \ll a_c$ so the 
dispersion is monotonic and
mostly linear. (b) $a$ just below $a_c$ so a pronounced minimum
is developed near $k=k_0$, but no instability yet. (c) $a > a_c$, so $\epsilon_k$
becomes negative for $k\approx k_0$, indicating edge reconstruction instability.
See text for details.
}
\label{dispersion}
\end{figure}

Once the gradient terms are included, the chiral boson dispersion is
no longer perfectly linear and has a {\em downward} curvature for small $k$:
\beq
\epsilon_k=v(k-ak^3+bk^5+\cdots),
\label{spectrum}
\eeq
as illustrated in Fig. (\ref{dispersion}) for different values of $a$;
such non-linear dispersion has been seen in our numerical studies\cite{specnote}.
In particular, there exist a critical
point $a_c=2\sqrt{b}$; at this point there is a momentum $k_0=b^{-1/4}$
at which
$\epsilon_{k_0}=0$. For $a > a_c$ we have $\epsilon_k < 0$ for $k\approx k_0$;
{\em i.e.}, the ground state of the system is no longer the vacuum of the
chiral bosons, and the chiral bosons will condense into states with
$k\approx k_0$! This instability is precisely the instability toward
edge reconstruction, as it leads to an increase of the momentum 
of the ground state, 
and edge density oscillation\cite{sondhi}. 
In this case we must include a repulsive interaction among the
chiral bosons to maintain the stability of the model\cite{internote}:
\beq
H_{int}=\int{dx}[u_3\rho^3(x)+u_4\rho^4(x)+\cdots].
\label{int}
\eeq
By stopping at the quartic order we are {\em assuming} that $u_4 > 0$.
Combining Eqs. (\ref{rch}) and (\ref{int}), we propose the following
field theory to describe the edge of a $\nu=1/m$ quantum Hall liquid:
\beqarr
S&=&{m\over 4\pi}\int{dtdx}\{\partial_t\phi\partial_x\phi
-v[(\partial_x\phi)^2-a(\partial^2_x\phi)^2+b(\partial^3_x\phi)^2]\}
\nonumber\\
&-&\int{dtdx}[u_3(\partial_x\phi)^3+u_4(\partial_x\phi)^4].
\label{action}
\eeqarr
Here $S$ is the action, and the real bosonic field $\phi(x)$ is related to the
edge electron density through $\rho(x)=\partial_x\phi(x)/(2\pi)$; in the
special case $a=b=u_3=u_4=0$ it reduces to the action of the
original CLL theory\cite{wen}. This model
supports two phases; for $a < a_c$ the ground state is the vacuum state of the
bosons, which properly describes the edge without reconstruction; the low-energy
excitations are the chiral bosonic modes at small $k$. For $a > a_c$
there is a finite density of bosons in the ground state, mostly occupying modes
with $k\approx k_0$. This is the phase with edge reconstruction. These 1D
bosons with repulsive interaction form an ordinary (or non-chiral)
Luttinger liquid which in turn can be
mapped onto non-chiral free bosons; thus in addition to the
chiral branch of bosons, there is also a non-chiral branch of low-energy bosonic
excitations in this case, which we have seen in our
numerical studies\cite{wan,wry}.

To study the critical behavior of this transition, we formally integrate out the
high energy modes in Eq. (\ref{action}), and focus on the low-energy modes near
$k\approx 0$ and $k\approx k_0$. These low-energy modes are conveniently
described in terms of the
following slowly-varying (in space) bosonic fields (time dependence is implicit
here):
\beqarr
\phi_1(x)&=&{1\over \sqrt{L}}\sum_{|k| < \Lambda}\phi_ke^{ikx};\\
\phi_2(x)&=&{1\over \sqrt{L}}\sum_{|k| < \Lambda}\phi_{k_0+k}e^{ikx}.
\eeqarr 
Here $L$ is the length of the edge, and
$\Lambda$ is a cutoff in momentum space. We note that while $\phi_1(x)$ is
a real field, $\phi_2(x)$ is actually a {\em complex} field. In terms of 
$\phi_1(x)$ and $\phi_2(x)$, the original action in Eq. (\ref{action}) takes 
the form:
\beq
S=S_1+S_2+S_{12},
\eeq
where
\beqarr
&S_1&={m\over 4\pi}\int{dtdx}[\partial_t\phi_1\partial_x\phi_1
-v(\partial_x\phi_1)^2];\\
&S_2&=\int{dtdx}(i\overline{\psi}_2\partial_t{\psi}_2
-{|\partial_x\psi_2|^2\over 2m^*}+\mu|\psi_2|^2-\tilde{u}|\psi_2|^4);\\
&S_{12}&=-\int{dtdx}[\tilde{u}_3(\partial_x\phi_1)|\psi_2|^2
+\tilde{u}_4(\partial_x\phi_1)^2|\psi_2|^2].
\eeqarr
Here $\psi_2=\sqrt{mk_0/ 2\pi}\phi^*_2$,
$\mu\propto a-a_c$, $1/m^*\approx 8vb^{1/4}$,
$\tilde{u}$, $\tilde{u}_3$ and $\tilde{u}_4$ are proportional to $u, u_3$ and
$u_4$ at tree level but receive loop 
renormalization from integrating out higher
energy modes, and we have neglected terms that scale to zero in the long-wave
length limit (like $(\partial_t\psi_2)(\partial_x\overline{\psi}_2)$,
$(\partial_x\phi_1)^3$, and $(\partial_x\phi_1)^4$ etc).
We see $S_1$ takes exactly the same form as the original CLL action
(but with a much reduced momentum cutoff), while
$S_2$ is identical to the action of 1D {\em non-relativistic}  
bosons with repulsive interaction\cite{sachdev}. 
If we neglect $S_{12}$ that describes 
interaction between $\phi_1(x)$ and $\psi_2(x)$ for the moment,
it is known\cite{sachdev} that the system undergoes a second order phase
transition from 
the vacuum of bosons (corresponding  to edge without reconstruction),
to a new ground state with a finite bosons density in it (corresponding to 
reconstructed edge), 
at the critical point $\mu=0$ (or $a=a_c$). The effect of $S_{12}$ may be 
taken into account by integrating out $\phi_1$ in $S$, which results in 
{\em finite} renormalization of $\tilde{u}$ in $S_2$.
Assuming that the renormalized value of $\tilde{u}$ to be positive ({\em i.e.},
the effective interaction of the bosons described 
by $\phi_2$ remains repulsive), neither the position of the critical point 
$\mu=0$ (or $a=a_c$), nor the critical property of the transition is changed
by the coupling between $\phi_1$ and $\phi_2$ as described by $S_{12}$.
The critical exponents of this transition are known exactly\cite{sachdev}:
$\nu=1/2$ and $z=2$, 
from which all other exponents can be deduced. In particular,
the boson density or the change of ground state momentum per unit length 
scales as: $n\sim\delta k\propto (d-d^*)^{1/2}$, where $d$ is a controlling
parameter (say, the distance between the dopant layer and the 2D electron
gas layer\cite{wan,wry}) that tunes the system through the transition, 
and $d^*$ is the critical point. 

On the other hand, if the renormalized quartic coupling turns 
out to be negative,
then the effective interaction between the bosons is {\em attractive}. In this
case higher order couplings need to be kept, and the transition between the 
two phases may become
first-order. Whether this is the case or not depends on the
details of edge confining potential and electron-electron interaction. 
Hartree-Fock study of edge reconstruction at bulk filling $\nu=1$ appears to 
suggest the transition is indeed first order in that case, for the type of 
confining potential that was used\cite{chamon}.

No matter the transition is first-order or second-order, the condensed 
non-chiral bosons described by $\psi_2$ in the reconstructed phase
form an ordinary or non-chiral Luttinger liquid. Perhaps the easiest way to 
obtain the Gaussian field theory (or Luttinger liquid) description of these
non-chiral bosons from Eqs. (9-11)
is to write $\psi_2$ as $\psi_2(x,t)=\sqrt{n(x,t)}\exp[i\varphi(x,t)]$,
and then integrate out the fluctuation of the boson density $n$ about its mean
value $\overline{n}\approx \mu/2\tilde{u}$ 
in $S$\cite{popov}, after which one obtains  
\beq
S=S_1+S_\varphi+S_{int},
\label{newac}
\eeq
where $S_1$ takes the same form as in Eq. (9) with renormalized velocity $v$,
and\cite{popov}
\beq
S_\varphi=\int{dtdx}{\overline{n}\over 2m^*}
\left[{1\over v_\varphi^2}(\partial_t 
\varphi)^2 -(\partial_x\varphi)^2\right], 
\eeq
and $v_\varphi\approx \sqrt{2\tilde{u}\overline{n}/m^*}\approx\sqrt{\mu/m^*}$.
Physically $\partial_t \varphi$ and $\partial_x\varphi$ are proportional to 
the density and current of the (non-chiral) bosons described by $\psi_2$, 
through the
Josephson relation. $S_{int}$ describes the interaction between the chiral 
and non-chiral bosons through density-density coupling:
\beq
S_{int}=-g\int{dtdx}(\partial_t \varphi)(\partial_x\phi_1),
\eeq
where $g\approx\tilde{u}_3/2\tilde{u}$.
Thus edge reconstruction
adds two more propagating edge modes in the edge spectrum, one propagating
in the forward direction and another in the {\em backward} direction; these new
modes are coupled to the original long wave-length chiral boson modes. 
Mathematically, the action of Eq. 
(\ref{newac}) is equivalent to that of a single CLL mode coupled to 
one-dimensional acoustic
phonons, a model that has been considered before in very different 
contexts\cite{eggert,rosenow}.

Obviously, multiple edge reconstruction transitions can occur, if there are
multiple local minima in the chiral boson spectrum Eq. (\ref{spectrum})
that go through
zero. The critical behavior of these additional transitions will be the same,
and each transition will introduce two more edge modes, propagating in opposite
directions.  

We now turn our discussion to the effect of edge reconstruction on single
electron Green's function. 
Within the CLL theory, the charge and statistics of the electron
operator uniquely determines its form in terms of the edge density field
$\phi(x)$ to be\cite{wen}
$\Psi(x)\propto e^{-im\phi(x)}.
$
We are interested
in the long-time or low-energy/frequency behavior of the electron Green's 
function, which is dominated by the low-energy modes of $\phi(x)$. For edges
without reconstruction they are the long-wave length modes of $\phi(x)$ with
$k\approx 0$, while in the presence of edge reconstruction they are modes
with $k\approx 0$ and $k\approx \pm k_0$. Thus in the latter case we write the
electron operator as  
\beqarr
&\Psi(x)&\sim \exp\{-im[\phi_1(x)+\phi_2(x)e^{ik_0x}+\phi^*_2(x)e^{-ik_0x}]\}
\nonumber\\
&\approx&\exp\{-im[\phi_1(x)+ce^{i(k_0x-\varphi(x))}+ce^{i(\varphi(x)-k_0x)}]\}
\nonumber\\
&=&e^{-im\phi_1(x)}\sum_{l=0}^{\infty}(-2imc)^l
\cos^l[\varphi(x)-k_0x].
\eeqarr
Here the constant $c\approx \sqrt{2\pi\overline{n}/mk_0}$. We have thus 
expressed the electron operator in terms of the Gaussian variables $\phi_1$ and
$\varphi$, whose correlation functions are controlled by the quadratic low-energy
effective action $S$ in Eq. (\ref{newac}). Thus the long-time behavior of the
electron Green's function can be determined straightforwardly:
\beq
G(x=0, t)=\langle\Psi(0, t)\Psi^\dagger(0, t=0)\rangle
=\sum_{l=-\infty}^{\infty}A_lt^{-\gamma_l},
\eeq
where $A_l$'s are some constants, and $\gamma_l$ is twice the scaling dimension
of the operator $O_l=e^{-im\phi_1+il\varphi}$: 
$\langle O_l(0,t)O_l(0,0)\rangle\sim t^{-\gamma_l}$. The minimum value of 
$\gamma_l$ controls Green's function in the long time limit, and thus
the $I-V$ characteristics of point contact tunneling 
between the edge and a Fermi liquid metal is
$I\sim V^\alpha$ with $\alpha=\gamma^{min}$.
Using a generalized Bogliubov transformation\cite{eggert} 
to obtain the eigen modes of
the action Eq. (\ref{newac}) and express $O_l$ as combination of the 
eigen modes, one can easily show that all $\gamma_l$'s are non-universal and
satisfy $\gamma_l > m$. Thus the tunneling exponent $\gamma^{min}$ is 
{\em non-universal}. This is a consequence of the lack of maximum chirality
due to edge reconstruction, and in sharp contrast with the case without edge
reconstruction, where the tunneling exponent $\alpha=m$ and 
is thus universal\cite{wen}.

In real samples, one expects $v$, the chiral charge mode velocity, to be much 
larger than all other velocity scales, especially 
the non-chiral mode velocity $v_\varphi$. This is because 
$v$ is controlled by the long-range Coulomb interaction and thus diverges as 
$\log k$ in the long-wave length limit; this divergence may be cut off by 
metallic gates placed near the sample, but only at very long length scales.
On the other hand $v_\varphi\approx\sqrt{\mu/m^*}$ is a neutral mode velocity,
and expected to be low since there is no reason for the chiral boson mode near
the instabilities to develop very deep minima and have large
curvature. 
In the limit of large $v$, we find
\beq
\alpha=\gamma_0=m[1+ v_{coup}v_\varphi/v^2+O(1/v^4)],
\eeq
where $v_{coup}\approx m\tilde{u}_3^2/\pi\tilde{u}$ is a positive 
velocity scale that 
parametrizes the strength of coupling between charge and neutral modes.
We thus find that the tunneling exponent is {\em increased}
by a small and non-universal
amount due to edge reconstruction. While consistent with experimental
findings that $\alpha\approx m$ for $m=3$, we note that in all experiments 
$\alpha$ is slight {\em below} 3 (typically by about $10\%$). 
This, however, may be due to the fact that 
electrostatic forces provided by nearby gates tend {\em increase} the  
electron density by $20-30\%$ over several hundred nanometers in the
edge region\cite{levitov01};
thus the the actual value of $\alpha$ that correspond to $\nu=1/3$ in the edge
region may very well be slightly above 3, consistent with our result.

Recent edge tunneling experiments motivated a considerable amount of 
theoretical 
work\cite{levitov01,conti98,lee98,lopez99,goldman}.
One of 
the main focuses of these studies is the apparent inverse relation between 
the tunneling exponent $\alpha$ and bulk filling factor $\nu$ observed in at
least one of the experiments\cite{grayson}: 
$\alpha\approx 1/\nu$. In one of the proposals put forward to explain this
approximate dependence, Lee and Wen\cite{lee98} made a key assumption
that there exist 
neutral mode(s) in the system whose velocity is extremely low. As discussed
above,
edge reconstruction can naturally lead to neutral modes with low velocities.
Thus edge reconstruction can not only explain the lack of universality in
$\alpha$ near
principle filling factors, but may also be a key ingredient 
in the understanding 
of its general dependence on filling factor.

The author benefited greatly from collaboration with Xin Wan and Ed Rezayi on
numerical studies of edge reconstruction.
He also thanks Claudio Chamon, Matthew Fisher, Bertrand Halperin, Chetan Nayak, 
Leo Radzihovsky,
Shivaji Sondhi,
and Xiao-Gang Wen for stimulating discussions, and Aspen Center
for Physics for hospitality where part of this work was performed.
This work was supported by
NSF grants No. DMR-9971541 and No. DMR-0225698,
and the A. P. Sloan Foundation.


\begin{references}

\bibitem{review} For a review, see C. L. Kane and M. P. A. Fisher, in
{\em Perspectives in Quantum Hall Effect},
edited by S. Das Sarma and A. Pinczuk (Wiley, New York, 1997).

\bibitem{wen} X.-G. Wen, Int. J. Mod. Phys. B {\bf 6}, 1711 (1992).

\bibitem{chang} A. M. Chang, L. N. Pfeiffer, and K. W. West, Phys. Rev. Lett.
{\bf 77}, 2538 (1996).

\bibitem{grayson} M. Grayson, D. C. Tsui, L. N. Pfeiffer, K. W. West, and
A. M. Chang, Phys. Rev. Lett. {\bf 80}, 1062
(1998).

\bibitem{chang1} A. M. Chang, M. K. Wu, C. C. Chi, L. N. Pfeiffer,
and K. W. West, Phys. Rev. Lett. {\bf 86}, 143
(2001).

\bibitem{hilke} M. Hilke, D. C. Tsui, M. Grayson, L. N. Pfeiffer,
and K. W. West, Phys. Rev. Lett. {\bf 87}, 186806 (2001).

\bibitem{macdonald}A. H. MacDonald, S. R. E. Yang, and M. D. Johnson,
Aus. J. Phys. {\bf 46}, 345 (1993).

\bibitem{chamon} C. Chamon and X.-G. Wen, Phys. Rev. B {\bf 49},
8227 (1994).

\bibitem{wan} X. Wan, K. Yang, and E. H. Rezayi, Phys. Rev. Lett. {\bf 88},
056802 (2002).

\bibitem{wry} X. Wan, E. H. Rezayi, and K. Yang, cond-mat/0302341.

\bibitem{note} Once the presence and pattern of the edge reconstruction is
known, such new modes can be introduced by hand according to the pattern of
reconstruction. See Ref.\onlinecite{chamon}.

\bibitem{specnote} See Fig. 4 of Ref. \onlinecite{wry}.

\bibitem{sondhi} In the case of a $\nu=1$ edge, the softening of edge modes at
finite wave vector that leads to edge reconstruction has been observed in 
time-dependent Hartree-Fock studies: A. Karlhede, K. Lejnell, and S. L. Sondhi,
Phys. Rev. B {\bf 60}, 15948 (1999); J. Sjostrand, A. Eklund, and A. Karlhede,
Phys. Rev. B {\bf 66}, 165308 (2002).

\bibitem{internote} Physically such interaction terms can come from,
{\em e.g.}, the non-linearity of the confining potential at the edge, in a way
similar to the non-linearity of fermion dispersion leading to interaction
terms in the bosonized Hamiltonian of ordinary Luttinger liquid; see,
{\em e.g.}, F. D. M. Haldane, J. Phys. C, {\bf 14}, 2585 (1981). 
While irrelevant
in the long-wave length limit for edges {\em without} reconstruction, we must
keep them here to maintain the stability of the theory when there is edge
reconstruction. 

\bibitem{sachdev} See, {\em e.g.}, S. Sachdev, {\em Quantum Phase Transitions},
Cambridge, 1999, Ch. 11.

\bibitem{popov} See, {\em e.g.}, V. N. Popov, {\em Functional Integrals and
Collective Excitations}, Cambridge, 1987.

\bibitem{eggert} O. Heinonen and S. Eggert, Phys. Rev. Lett. {\bf 77}, 358 (1996).

\bibitem{rosenow} B. Rosenow and B. I. Halperin, Phys. Rev. Lett. {\bf 88},
096404 (2002).

\bibitem{levitov01} L. S. Levitov, A. V. Shytov, and
B. I. Halperin, Phys. Rev. B {\bf 64}, 075322 (2001).

\bibitem{conti98} S. Conti and G. Vignale, J. Phys. Condens. Matt.
{\bf 10}, L779 (1998);
J.H. Han and D.J. Thouless, Phys. Rev. B
{\bf 55}, 1926 (1997);
U. Z\"ulicke and A.H. MacDonald, Phys. Rev. B
{\bf 60}, 1836 (1999); U. Z\"ulicke, J. J. Palacios,
 and A.H. MacDonald, Phys. Rev. B
{\bf 67}, 045303 (2003).

\bibitem{lee98} D.-H. Lee and X.-G. Wen, cond-mat/9809160.

\bibitem{lopez99} A. Lopez and E. Fradkin, Phys. Rev. B
{\bf 59}, 15323 (1999).

\bibitem{goldman} V. J. Goldman and E. V. Tsiper, Phys. Rev. Lett. {\bf 86}, 
5841 (2001); E. V. Tsiper and V. J. Goldman, Phys. Rev. B {\bf 64}, 
165311 (2001);
S. S. Mandal and J. K. Jain,
Phys. Rev. Lett. {\bf 89}, 096801 (2002).

\end{references}
\end{document}